\RequirePackage[switch,columnwise]{lineno}

\documentclass[amsmath, amssymb, amsthm, amsfonts, superscriptaddress, reprint, aip, apl]{revtex4-1}

\usepackage{graphicx}
\usepackage{mathptmx}
\usepackage{color}
\usepackage{soul}

\usepackage{hyperref}
\hypersetup{colorlinks} 
\hypersetup{
    linkcolor=blue, 
    citecolor=blue,
    filecolor=magenta, 
    urlcolor=blue}

\begin{document}

\title{Anisotropic structural dynamics of monolayer crystals revealed by
femtosecond surface x-ray scattering}

\author{I-Cheng Tung}
\affiliation{Advanced Photon Source, Argonne National Laboratory, Argonne, IL 60439, USA}

\author{Aravind Krishnamoorthy}
\affiliation{Collaboratory of Advanced Computing and Simulations, University of Southern California, Los Angeles, CA 90089, USA}

\author{Sridhar Sadasivam}
\affiliation{Center for Nanoscale Materials, Argonne National Laboratory, Argonne, IL 60439, USA}

\author{Hua Zhou}
\affiliation{Advanced Photon Source, Argonne National Laboratory, Argonne, IL 60439, USA}

\author{Qi Zhang}
\affiliation{Advanced Photon Source, Argonne National Laboratory, Argonne, IL 60439, USA}

\author{Kyle L. Seyler}
\affiliation{Department of Physics, University of Washington, Seattle, WA 98195, USA}

\author{Genevieve Clark}
\affiliation{Department of Physics, University of Washington, Seattle, WA 98195, USA}

\author{Ehren M. Mannebach}
\affiliation{Department of Materials Science and Engineering, Stanford University, Stanford, CA 94305, USA}

\author{Clara Nyby}
\affiliation{Department of Chemistry, Stanford University, Stanford, CA 94305, USA}

\author{Friederike Ernst}
\affiliation{Department of Applied Physics, Stanford University, Stanford, CA 94305}
\affiliation{PULSE Institute, SLAC National Accelerator Laboratory, Menlo Park, CA 94025, USA}

\author{Diling Zhu}
\affiliation{SLAC National Accelerator Laboratory, Menlo Park, CA 94025, USA}

\author{James M. Glownia}
\affiliation{SLAC National Accelerator Laboratory, Menlo Park, CA 94025, USA}

\author{Michael E. Kozina}
\affiliation{SLAC National Accelerator Laboratory, Menlo Park, CA 94025, USA}

\author{Sanghoon Song}
\affiliation{SLAC National Accelerator Laboratory, Menlo Park, CA 94025, USA}

\author{Silke Nelson}
\affiliation{SLAC National Accelerator Laboratory, Menlo Park, CA 94025, USA}

\author{Hiroyuki Kumazoe}
\affiliation{Department of Physics, Kumamoto University, Kumamoto 860-8555, Japan}

\author{Fuyuki Shimojo}
\affiliation{Department of Physics, Kumamoto University, Kumamoto 860-8555, Japan}

\author{Rajiv K. Kalia}
\affiliation{Collaboratory of Advanced Computing and Simulations, University of Southern California, Los Angeles, CA 90089, USA}

\author{Priya Vashishta}
\affiliation{Collaboratory of Advanced Computing and Simulations, University of Southern California, Los Angeles, CA 90089, USA}

\author{Pierre Darancet}
\affiliation{Center for Nanoscale Materials, Argonne National Laboratory, Argonne, IL 60439, USA}

\author{Tony F. Heinz}
\affiliation{Department of Applied Physics, Stanford University, Stanford, CA 94305, USA}
\affiliation{PULSE Institute, SLAC National Accelerator Laboratory, Menlo Park, CA 94025, USA}
\affiliation{Stanford Institute for Materials and Energy Sciences, SLAC National Accelerator Laboratory, Menlo Park, CA 94025, USA}

\author{Aiichiro Nakano}
\affiliation{Collaboratory of Advanced Computing and Simulations, University of Southern California, Los Angeles, CA 90089, USA}

\author{Xiaodong Xu}
\affiliation{Department of Physics, University of Washington, Seattle, WA 98195, USA}

\author{Aaron M. Lindenberg} 
\affiliation{Department of Materials Science and Engineering, Stanford University, Stanford, CA 94305, USA}
\affiliation{PULSE Institute, SLAC National Accelerator Laboratory, Menlo Park, CA 94025, USA}
\affiliation{Stanford Institute for Materials and Energy Sciences, SLAC National Accelerator Laboratory, Menlo Park, CA 94025, USA}

\author{Haidan Wen}
\email{wen@anl.gov}
\affiliation{Advanced Photon Source, Argonne National Laboratory, Argonne, IL 60439, USA}

\date{\today}
\begin{abstract}
X-ray scattering is one of the primary tools to determine crystallographic configuration with atomic accuracy. However, the measurement of ultrafast structural dynamics in monolayer crystals remains a long-standing challenge due to a significant reduction of diffraction volume and complexity of data analysis, prohibiting the application of ultrafast x-ray scattering to study nonequilibrium structural properties at the two-dimensional limit. Here, we demonstrate femtosecond surface x-ray diffraction in combination with crystallographic model-refinement calculations to quantify the ultrafast structural dynamics of monolayer WSe$_2$ crystals supported on a substrate. We found the absorbed optical photon energy is preferably coupled to the in-plane lattice vibrations within 2 picoseconds while the out-of-plane lattice vibration amplitude remains unchanged during the first 10 picoseconds. The model-assisted fitting suggests an asymmetric intralayer spacing change upon excitation. The observed nonequilibrium anisotropic structural dynamics in two-dimensional materials agrees with first-principles nonadiabatic modeling in both real and momentum space, marking the distinct structural dynamics of monolayer crystals from their bulk counterparts. The demonstrated methods unlock the benefit of surface sensitive x-ray scattering to quantitatively measure ultrafast structural dynamics in atomically thin materials and across interfaces.
\end{abstract}

\maketitle

The development of van der Waals (vdW) materials has opened up possibilities for the exploration of new physics in the two-dimensional (2D) limit \cite{Novoselov:2016ik,Ajayan:2016cd}. Strong light-matter interaction in 2D materials allows optical control of electronic, spin and valley degrees of freedom, in which the structure of 2D materials is usually approximated to be stationary with weak optical excitations. With increasing optical excitation strengths, nonlinear processes start to occur \cite{Chernikov:2015gc,langer_lightwave-driven_2016,Rivera:2016ie} and the Born-Oppenheimer approximation may not be applicable. In addition, toward the monolayer limit, the influence of the surrounding environment on the properties of 2D systems becomes increasingly important \cite{Sun:2017hf,Nika:2012jv,frigge2017,Sun:2017hf}. For example, the anomalously large thermal conductivity of graphene may find applications in efficient thermal removal \cite{Balandin:2008fv,Ghosh:2010ja}. Unconventional interface superconductivity is ascribed to the unique electron-phonon coupling between the FeSe and the SrTiO$_3$ interface \cite{Lee:2014bw}, and unusual exciton-phonon interactions have been observed across the interface of vdW heterostructures as well as between monolayer semiconductors and crystalline substrates \cite{Chow:2017jy,jin_interlayer_2017}. However, in contrast to many investigations that discover unique electronic, spin, thermal, and optical properties of 2D materials and their interfaces, limited knowledge of the associated structural dynamics has been obtained. Understanding of these emergent phenomena in 2D systems beyond thin films \cite{gerber_femtosecond_2017} calls for a direct quantitative measurement of the lattice dynamics of monolayer crystals at and across crystalline interfaces on ultrafast time scales.

\begin{figure*}[t]
\centering
\includegraphics[width=1\textwidth]{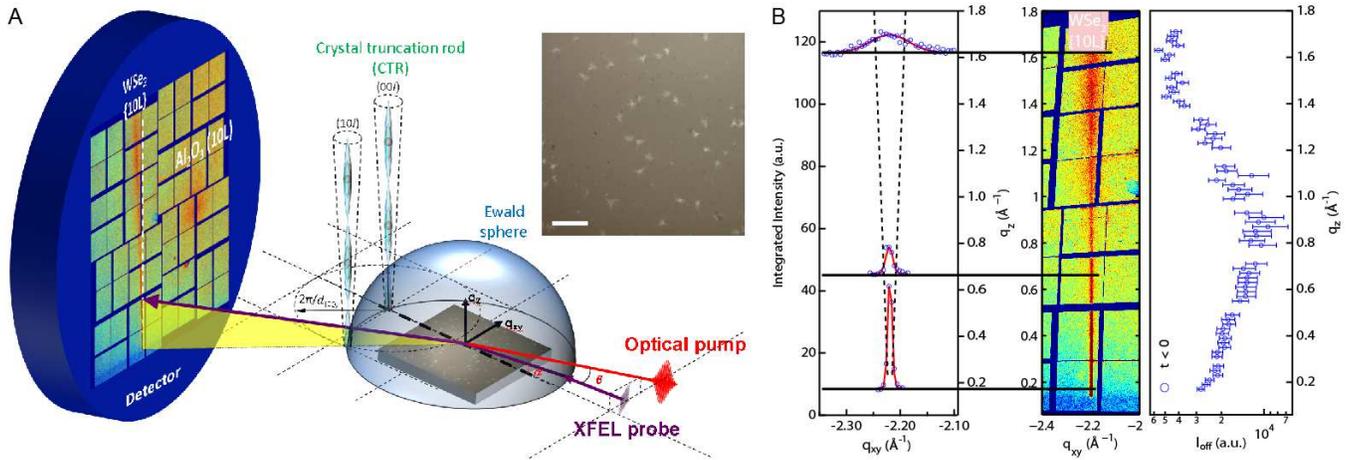}
\caption{
(A) A schematic diagram of femtosecond x-ray diffraction. The specular $(00l)$ and the $(10l)$ rods are shown representatively across the Ewald sphere. The X-ray probe beam strike on the sample close to grazing incidence ($\alpha = 1^{\circ}$) and the optical pump beam is overlap with the x-ray beam with a horizontal crossing angle $\beta = 5^{\circ}$. The inset shows the optical images of the sample surface. Scale bar is 30 $\mu$m. (B) (center) Corrected reciprocal space map around the off-specular $\{10l\}$ rod. (left) Rod profiles for different $q_z$ (blue spots) and Gaussian fits (red), with an offset that corresponds to the $q_z$ values. The dashed lines that connect the curves at approximately their FWHM is drawn as a guide to the eye. (right) Integrated intensity values for the ground state $\{10l\}$ CTR, obtained by integration at different out-of-plane momentum transfer values $q_z$.
}
\label{fig:measurement_setup}
\end{figure*}

 With advances in generating ultrabright and ultrashort electron and x-ray pulses, tracking atomic structural dynamics in 2D materials on femtosecond time scales with high precision becomes possible. For example, ultrafast electron diffraction (UED) and microscopy with large scattering cross section allows measurements of transient structural changes in various multilayer systems \cite{Raman:2008hb,Lin:2017js,waldecker:2017prl,cremons_femtosecond_2016,cremons_defect-mediated_2017} and at surfaces \cite{ruan_ultrafast_PNAS2004,ruan_ultrafast_science_2004,frigge2017}.  At the monolayer limit, UED in a transmission geometry has been used to characterize in-plane structural dynamics of monolayer 2D materials \cite{Mannebach:2015de,Hu:2016em}, while the out-of-plane structural dynamics of a monolayer crystal was only inferred \cite{Hu:2016em}. Comparing with ultrafast electron diffraction, ultrafast x-ray diffraction has been an indispensable tool for studying structural dynamics in thin films, providing evidence on non-thermal structural response in transition metal dichacogenides \cite{mannebach_dynamic_2017} and electron-phonon coupling in superconducting FeSe \cite{gerber_femtosecond_2017}. In particular, ultrafast surface x-ray surface scattering can provide direct and quantitative measurements of structural changes by recording diffraction profile with non-zero momentum transfer along the out-of-plane direction, in contrast to techniques that only measure in-plane diffraction peaks or surface sensitive optical probes close to Brillouin zone center \cite{shen_phase-sensitive_2013}. But the direct structural characterization of non-equilibrium processes within a monolayer crystal remains a challenge due to significantly reduced scattering volume. Ultrafast three-dimensional atomic-scale rearrangements of monolayer crystals on supporting substrates, the most common platform of 2D phenomena and configuration of 2D devices, have not yet been mapped out.

In this article, we report the first femtosecond surface X-ray diffraction (fSXD) study of monolayer crystals, which is enabled by the ultrahigh single-pulse brightness and ultrashort pulse durations of hard x-ray radiation of free electron lasers at the Linac Coherent Light Source (LCLS). Monolayer WSe$_2$ is chosen as the model system because it hosts strong exciton-phonon coupling with substantial impact in valley exciton dynamics\cite{jones_excitonic_2016,mishra_exciton-phonon_2018}, ultrafast dynamics \cite{poellmann_resonant_2015,langer_lightwave-driven_2016}, unusual exciton- interfacial phonon interactions at vdW interface\cite{Chow:2017jy,jin_interlayer_2017}, and unique chiral phonons\cite{zhu_observation_2018}, while its x-ray scattering cross section is relatively large in the family of transition metal dichalcogenides. In combination with crystallographic model-refinement calculations, we capture the lattice motions of monolayer WSe$_2$ along both in-plane and out-of-plane directions upon optical excitation. In particular, we find direct structural evidence of an anisotropic energy relaxation pathway that favors electron-phonon coupling along the in-plane direction occurring on the time scale of 2 ps, subsequently followed by an in-plane lattice expansion. On the contrary, the out-of-plane lattice vibration remains unchanged within the measurement window of 10 ps. The strongly anisotropic response agrees with first-principles simulation in real and momentum spaces, illuminating a key process of anisotropic nonequilibrium lattice response in 2D materials. In addition, the model assisted fitting suggests that the intralayer spacing change in monolayers are asymmetric, in contrast to a symmetric intralayer compression of bulk WSe$_2$ on ps time scales, underlying a distinct structural response of monolayer crystals on a supporting substrate. The demonstration of fSXD opens opportunities to study nonequilibrium physics of monolayer crystals at interfaces with atomic resolution on ultrafast time scales.

\begin{figure}[t]
\centering
\includegraphics[width=0.45\textwidth]{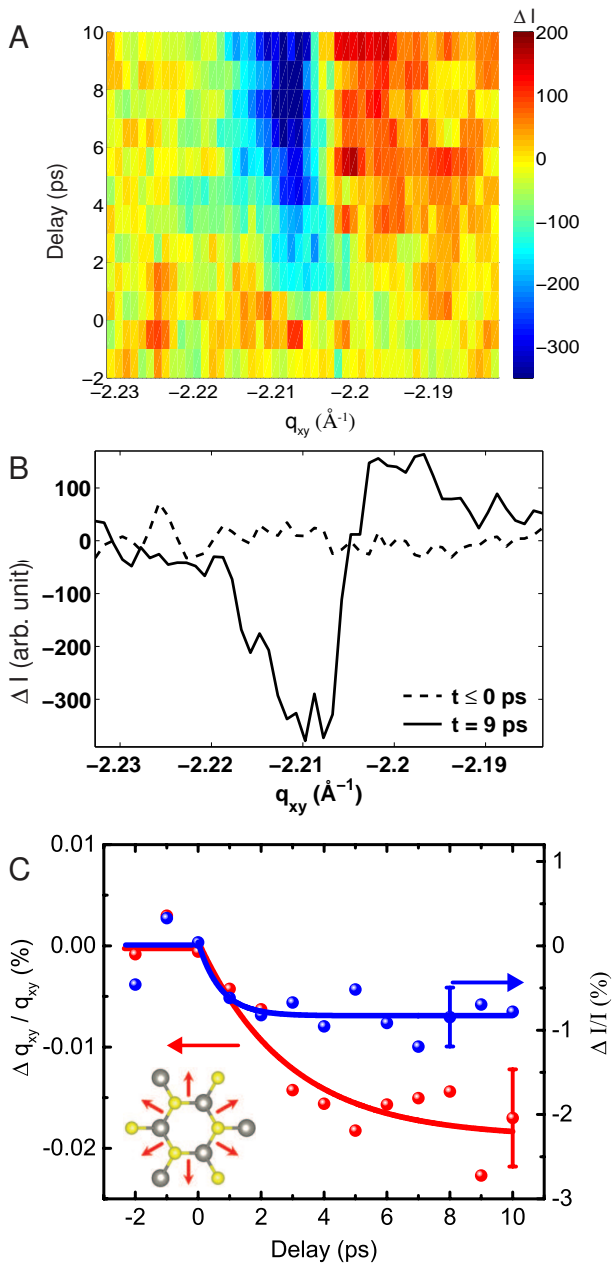}
\caption{
(A) Time-dependent differential diffraction intensity as projected to $q_{xy}$ axis integrated in the range $q_{z}=0.16-0.5$\ \AA$^{-1}$. (B) The differential diffraction intensity profile as a function of $q_{xy}$ before and after optical excitation. (C) Time-dependent in-plane rod position shift $\Delta q/q_0$ and diffraction intensity change $\Delta I/I_0$ of the $\{10L\}$ reflection. The solid lines are fits presented in the text. The error bars are the fitting errors. 
}
\label{fig:inplane_dynamics}
\end{figure}

\subsection*{In-plane Structural Dynamics}
We first characterize the sample at steady state by surface x-ray scattering (Fig.~\ref{fig:measurement_setup}A). The samples are WSe$_2$ monolayer crystals with an averaged flake size of a few $\mu m$ homogeneously covering the Al$_2$O$_3$ substrate (See Methods). Since the crystal dimension is highly confined along the out-of-plane direction, its Fourier transformation in the reciprocal space shows as a vertical streak pattern along the crystal truncation rod (CTR) of a both-side "truncated" crystal \cite{robinson_surface_1992} (See Methods). The in-plane random orientations of WSe$_2$ flakes permit recording the off-specular $\{10l\}$ crystal truncation rod (CTR), the sum of $(10l)$ and $(01l)$ rods, on the area detector without rotating the sample, simultaneously measuring the diffraction intensity distribution along the in-plane ($q_{xy}$) and out-of-plane ($q_z$) direction in reciprocal space. The in-plane intensity profile of $\{10l\}$ rod is well fitted by a Gaussian shape with width increasing at higher $q_z$, indicating rippling of the monolayer crystals (Fig.~\ref{fig:measurement_setup}B) \cite{Brivio:2011ch,Meyer:2007kq}. 

The structural dynamics is measured by surface x-ray diffraction after optical excitation with 650 nm light using pump-probe technique (See Methods). To study the in-plane dynamics, we integrate the diffraction intensity along $q_{z}$ in the range of $0.16-0.5\ \AA^{-1}$ (Fig.~\ref{fig:measurement_setup}B) to obtain the diffraction intensity profile projected onto the $q_{xy}$ axis. Time-dependent intensity profiles are plotted as a function of delay in Fig.~\ref{fig:inplane_dynamics}A. The asymmetric wave-like differential profile after time zero (Fig.~\ref{fig:inplane_dynamics}B) indicates a diffraction intensity reduction and a rod position shift to lower $q_{xy}$ axis after excitation. 

The $\{10l\}$ Bragg rod position and intensity can be extracted as a function of time (Fig.~\ref{fig:inplane_dynamics}C). Following optical excitation, we observe a decrease in diffracted intensity as a result of Debye-Waller (DW) effect that can be fit to an exponential function $\Delta I(t) \sim A[1-\exp(-t/\tau)]$ with amplitude $A$ and time constant $\tau$ (blue curves, Fig.~\ref{fig:inplane_dynamics}C). The in-plane electron-phonon coupling time $\tau=0.8 \pm 0.4$ ps is faster than $1.83 \pm 0.13$ ps in bulk WSe$_2$ \cite{waldecker:2017prl}. The slowing down of energy coupling to the in-plane lattice vibration in bulk may have multiple origins, including lower defect density and indirect band gap in bulk, all of which can reduce electron-phonon scattering rates and thus slow down the energy transfer from electrons to in-plane lattice motion. The magnitude of $0.9\%$ reduction of the diffraction intensity of $\{10l\}$ rods corresponds to $\Delta \langle u_{xy}^2 \rangle=[\ln (I_{0}/I)]/q_{xy}^2=0.0018\ \AA^2$, where $I_0$ and $I$ are the diffraction intensity before and after the excitation \cite{Lin:2017js}, and $\langle u_{xy} \rangle$ represents the root mean square displacement (RMSD) along either $x$ or $y$ directions, assuming isotropic in-plane atomic displacement. The increase of $\langle u_{xy}^2 \rangle$ corresponds to an in-plane lattice temperature increase of $88$ K, consistent with the estimated temperature rise from the absorbed energy (Supplementary Note 3). The time-dependent change of the CTR position was determined by fitting a 1D Gaussian profile to the $\{10l\}$ rod intensity as projected to $q_{xy}$ axis. The peak position shifts as a function of time can be described by an one-exponential function $\Delta q(t) \sim A[1-\exp(-t_q/\tau_1)]$ (red curve, Fig.~\ref{fig:inplane_dynamics}C). The time constant of $\tau_1=2.9 \pm 1.2$ ps shows characteristic time of the in-plane lattice expansion, as a result of an in-plane acoustic wave propagation \cite{Hu:2016em,Mannebach:2015de}. Comparing with the freestanding multilayer samples \cite{cremons_femtosecond_2016,cremons_defect-mediated_2017}, the lack of oscillatory Bragg peak shift suggests an overdamped mode that may result from the interaction with the substrate.

\subsection*{Out-of-plane Structural Dynamics}

\begin{figure}[t]
\centering
\includegraphics[width=0.31\textwidth]{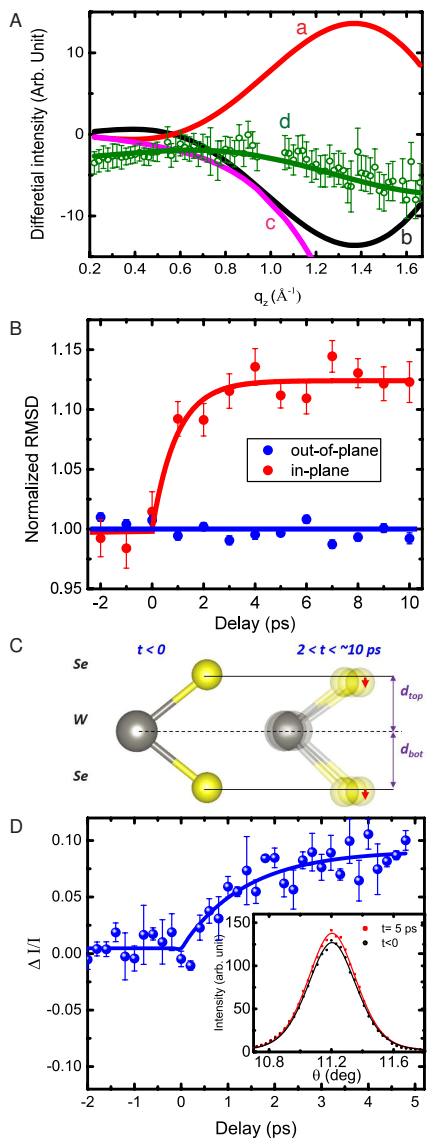}
\caption{
(A) Differential x-ray scattering intensity profile of WSe$_2$ monolayers along the $\{10l\}$ rod for $5 \leq \Delta t \leq 10$ ps. The solid lines are model calculations discussed in the text. (B) Normalized in-plane and out-of-plane RMSD of WSe$_2$ monolayers. The error bars are the standard deviation of the fitting parameters in all fitting trials. Solid curves are the guides to the eye. (C) Schematic diagram of the atomic structure model of monolayer WSe$_2$. The blurriness of atom positions indicates the increase of the in-plane RMSD. The red arrows show the possible intralayer distance changes. (D) The diffraction intensity change of 004 Bragg peak of a bulk WSe$_2$ crystal upon optical excitation. The error bars show the standard errors of consecutive scans. The inset shows the rocking curve with fits before and 5 ps after optical excitation. 
}
\label{fig:outofplane_dynamics}
\end{figure}

The out-of-plane lattice dynamics is obtained by comparing the measured CTRs with crystallographic model-refinement calculations of the diffraction intensity change along the $q_z$ as a function of time. The ground-state $\{10l\}$ CTR is shown in (Fig.~\ref{fig:measurement_setup}B). By subtracting the ground-state CTR from the excited-state CTR recorded at each delay, the time-dependent differential CTR along the $\{10l\}$ rod can be obtained. A representative differential CTR is shown as circles in Fig.~\ref{fig:outofplane_dynamics}A. The differential intensity is below zero, with a minimum change around $q_z=0.7$\ \AA$^{-1}$, indicating a loss of total diffraction intensity as a result of the increase of incoherent lattice vibrations. To qualitatively understand the $q_z$ dependent differential intensity profile, we compare the measured differenfial CTRs with calculated ones with structural parameters including RMSDs of atoms and intralayer spacings. The representative differential CTRs from four scenarios are shown in Fig.~\ref{fig:outofplane_dynamics}A: the intralayer spacing expands (a) or contracts (b) by $0.2\%$ without RMSD increase; the out-of-plane (c) and in-plane (d) RMSD increase of $12\%$ without d-spacing change. It is clear that only scenario d) matches the data well, indicating that the energy relaxation is preferable along the in-plane direction and symmetric intralayer d-spacing change does not agree with the measurement. The magnitude of RMSD change and intralayer spacing change as used in this simulation will be justified later. A systematical study of differential CTRs as a function of various structural parameters can be found in Supplementary Note 2.

\begin{figure*}[t]
\centering
\includegraphics[width=0.95\textwidth]{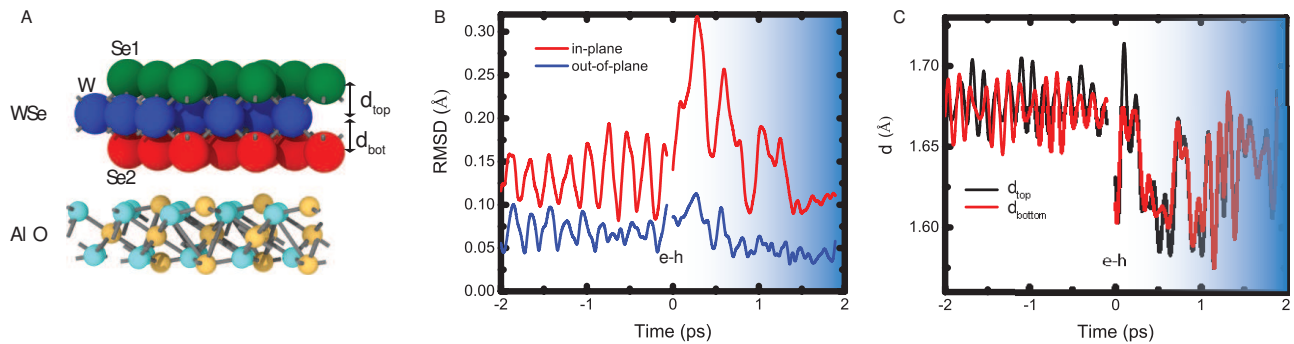}
\caption{
(A) The schematic of simulated system: WSe$_2$ monolayer supported on an Al$_2$O$_3$ substrate. (B) The in-plane and out-of-plane root mean squared displacements of all atoms  prior to and after electronic excitation respectively. "e-h" represents electron-hole injection at time zero with carrier density of $2.6\times10^{14}$ cm$^{-2}$. The blue region represents the coupling of the monolayer with the thermostat. (C) The intralayer spacing of W-Se atomic planes before and after electronic excitation.
}
\label{fig:theory}
\end{figure*}

To quantify the RMSD increase as a function of time, least-squares fitting was performed on the differential CTRs with the monolayer WSe$_2$ structure parameters allowed to vary (See Supplementary Note 1,2). Differential CTRs were fit with two parameters, the percentage change of in-plane and out-of-plane RMSD ($\sqrt{\langle u^2 \rangle/\langle u_0^2 \rangle}$), where $\sqrt{\langle u^2 \rangle}$ is the measured RMSD of the transient structure, and $\sqrt{\langle u_0^2 \rangle}$ is the calculated value for the ground state (See Supplementary Note 4). The intralayer spacing change is neglected for simplicity and will be treated in more details in the Discussion section. From the model-based analysis of the x-ray scattering intensity profiles along $\{10l\}$, we experimentally determined the corresponding mean square displacements shown in Fig.~\ref{fig:outofplane_dynamics}B. Consistent with the qualitative analysis, the out-of-plane RMSD remains almost the same within 10 ps, while the in-plane RMSD increase $12\%$, corresponds to $0.007\ \AA$ increase of the in-plane RMSD. The rise time is consistent with the blue curve in Fig.~\ref{fig:inplane_dynamics}C, as expected. This anomalous structural response is suggestive of a strong anisotropy of energy relaxation pathways.   

\subsection*{First principles simulations}

To give insight into the observed anisotropic energy relaxation pathway, we first perform nonadiabatic quantum molecular dynamics (NAQMD) simulations using the model shown in Fig.~\ref{fig:theory}A. NAQMD follows the trajectories of all atoms, while describing electronic excitations and nonadiabatic transitions between excited electronic states assisted by atomic motions based on time-dependent density functional theory and surface-hopping approaches, which enables it to describe photoexcitation dynamics involving coupled motion of electrons and nuclei \cite{Shimojo:2013hu}.

Electronic excitation is simulated by the instantaneous creation of electron-hole pairs in the monolayer-substrate simulation cell at time zero. Fig.~\ref{fig:theory}B show the calculated RMSD averaged over all atoms in the WSe$_2$ monolayer as a function of time at a simulated electron-hole concentration of 0 cm$^{-2}$ (i.e. no excitation) and $2.6\times10^{14}$ cm$^{-2}$ respectively. While the out-of-plane RMSD remains close to 0.08 $\AA$ independent of excitation, the in-plane RMSD increases strongly to over 0.3 $\AA$ approximately 300 fs after instantaneous electronic excitation. The magnitude of the in-plane RMSD increase in the simulation normalize to the electron density is in agreement of the measured value since the excitation strength of simulated e-h density $2.6\times10^{14}$ cm$^{-2}$ is twenty times higher than that of $1.3 \times 10^{13}$ cm$^{-2}$ used in the experiments. We note that the in-plane lattice response relaxes to the ground state within 1.5 ps due to the coupling of the WSe$_2$ monolayer to the Nose-Hoover thermostat at 300K. The use of Nose-Hoover thermostat is required to stabilize the MD simulation. The coupling with the thermostat unavoidably exaggerates the energy relaxation rate so that the relaxation of RMSD occurs on an artificially fast time scale. 

We note that the lateral size of WSe$_2$ crystal of $<$ 1 nm in the NAQMD simulation is much smaller than the rippling domain size of ~10 nm so that the nanoscale rippling effects are not taken into account. Complementary to NAQMD simulations, we track phonon populations of all vibrational modes as a function of time by calculating electron-phonon and third-order phonon-phonon interactions\cite{sadasivam_theory_2017} (Supplementary Note 5). The rippling effect is captured in these analytical calculations through the inclusion of occupation of out-of-plane phonon modes, as shown by the larger out-of-plane RMSD than the in-plane RMSD at the group state (Supplementary Note 4). The excitation is simulated by a sudden electronic temperature jump at time zero from 300 to 3000 K. The results shows a fast and significant increase of in-plane RMSD within 1 ps but a slow and small rise of RMSD within 10 ps (Supplementary Figure 2), consistent with the NAQMD simulations. The preferred energy relaxation along the in-plane direction at short times after excitation can be understood as the reduced density of out-of-plane phonon states in monolayer crystals\cite{ataca_comparative_2011}.

\section*{DISCUSSIONS}
We now discuss the change of the intralayer spacing between W and Se atomic planes within the monolayer upon optical excitation. We first present the bulk response as a reference. The intralayer dynamics of an exfoliated, 50 nm-thick WSe$_2$ crystal was measured by monitoring 004 Bragg peak upon excitation of the same 650 nm light pulses with an incident fluence of 9 mJ cm$^{-2}$. We found the diffraction peak intensity increases while the peak position does not shift (Fig.~\ref{fig:outofplane_dynamics}D), indicating a dominant atomic motion within unit cell within 5 ps, rather than unit cell size change that occurs at later time due to strain wave propagation \cite{mannebach_dynamic_2017}, or the tilting of the sample. By calculating the structure factor as a function of the intralayer spacing, the diffraction intensity change of 8$\%$ corresponds to an intralayer contraction of 1.4$\%$, which is 0.02 $\AA$ decrease of lattice spacing of W-Se layers. The intralayer contraction agrees with the NAQMD simulation shown in Fig.~\ref{fig:theory}C. Assuming a linear dependence of the d-spacing change on the pump fluence, we estimate a symmetric intralayer spacing change of 0.2$\%$ with the pump fluence (1.5 mJ cm$^{-2}$) used in the monolayer study. 

The model-assisted fitting of differential CTR of monolayer crystals can yield the intralayer spacings between W-Se layers if they are set as fitting parameters during the fitting procedure. However, the fitting cannot converge for non-zero symmetric intralayer spacing change that increase or decrease the same amount. We then consider to model the intralayer spacing independently as $d_{bot}$ and $d_{top}$, referring to the spacing between the top or bottom of Se layers to the middle W layer (Fig.~\ref{fig:outofplane_dynamics}C). Based on the intralayer change in the bulk, we constrain $\Delta d_{bot}$= 0.2$\%$. The fitting procedure converges and yields $\Delta d_{top}$=-0.25$\pm0.02\%$. The systematical studies of the influence of structural parameters on differential CTRs show that the asymmetric d-spacing change is required to fit the data, when considering non-zero intralayer spacing change (Supplementary Note 2). This indicates that the asymmetric infrared active mode $A^{''}_2$ is dominantly excited in monolayer crystals in contrast to the excitation of the $A_{1g}$ mode in the bulk\cite{ataca_comparative_2011}, which may be attributed to the breaking of the mirror symmetry respect to W layers due to the influence of substrate. The attempt to take account the substrate influence in the NAQMD simulation by introducing one unit-cell-thick Al$_2$O$_3$ under WSe$_2$ (Fig.~\ref{fig:theory}A) does not show a significant asymmetric change of intralayer spacing (Fig.~\ref{fig:theory}C), in comparison with the observed opposite sign change of intralayer spacing. This is potentially due to the one-unit-cell-thick substrate in the model. Simulations containing thicker substrates are computationally challenging and beyond the scope of this work.

\section*{Conclusions}
We demonstrated the femtosecond surface x-ray diffraction to study nonequilibrium structural dynamics of monolayer crystals on a supporting substrate. Upon optical excitation, we characterized both in-plane and out-of-plane atomic motions in monolayer WSe$_2$ crystals. We found that energy relaxation in lattice occurs anisotropically, favoring in-plane over out-of-plane direction. The corresponding intralyer spacing change in monolayer crystals are asymmetric, with Se atoms moving in the same direction respect to W atoms, potentially due to the influence of the substrate, and is in contrast to the symmetric intralayer lattice contraction in bulk crystals. These measurements enable the direct comparison to nonadiabatic quantum molecular dynamics simulations which largely reproduce the observations. Our general understanding of ultrafast optoelectronic processes under intense optical excitation need to be reevaluated to take into account the lattice degrees of freedom. The demonstrated femtosecond surface x-ray diffraction advances x-ray scattering technique for surface science on ultrafast time scales and opens new opportunities for a direct characterization of lattice structure at and across the interface of a wide range of low-dimensional material systems.

\section*{Methods}

\subsection*{Sample Preparation and static surface x-ray diffraction}
Monolayer WSe$_2$ films used in this work were grown using a physical vapor transport method \cite{Clark:2014fn} with the monolayer coverage over $95\%$ (grey color, Fig.~\ref{fig:measurement_setup}A inset) with a few mulilayer flakes $<3$ nm-thick (white color, Fig.~\ref{fig:measurement_setup}A inset). The small contribution from multilayer crystals does not alter CTR profile and thus was not included in the model-assisted data analysis. These crystals have an isotropic in-plane crystallite orientation distribution. Figure~\ref{fig:measurement_setup}A shows a schematic of the collinear pump-probe geometry used in this study. A typical grazing-incidence diffraction pattern from monolayer WSe$_2$ on a sapphire substrate is shown on the flat area detector as a tilted streaks in Fig.~\ref{fig:measurement_setup}A. A data correction procedure \cite{Jiang:2015fn} is applied to obtain diffraction intensity distribution in reciprocal space (Fig.~\ref{fig:measurement_setup}B). Since the in-plane lattice constant of WSe$_2$ is different from Al$_2$O$_3$ substrate, the monitored off-specular $\{10l\}$ CTR has no contribution from the sapphire substrate, which allows independent analysis of the monolayer structural properties without interference with the substrate \cite{Renaud:1998wt}. The $\{10l\}$ CTR of monolayer can then be discretely mapped by integrating the corrected diffraction pattern (Fig.~\ref{fig:measurement_setup}B) at different $q_z$ values \cite{Mannsfeld:2009iy}.

\subsection*{Femtosecond surface X-ray Diffraction}

Femtosecond x-ray scattering measurement were performed in grazing incident geometry at the X-ray Pump-Probe (XPP) instrument at the Linac Coherent Light Source (LCLS). The samples were kept under helium atmosphere during measurements. The monochromatic X-ray beam with a photon energy of 9.55 keV and 50 fs pulse duration strikes the sample surface at a grazing angle, $\alpha=1^{\circ}$ shown in Fig.~\ref{fig:measurement_setup}A. The large x-ray beam footprint resulting from grazing incidence geometry also serves to spread the x-ray beam with the cross section full-width-half-maxium of 50 $\times$ 50 $\mu m$ to minimize the sample degradation. The diffracted intensity is then detected by a 2D area detector, Cornell-SLAC Pixel Array Detector (CSPAD). The distance between sample and detector is 431 mm, calibrated by LaB6 diffraction pattern. The linearly polarized pump light pulses is derived from optical parametric amplifier with the wavelength of 650 nm and 200 $\times$ 200 $\mu m$, close to the absorption peak of the B-exciton of monolayer WSe$_2$ \cite{Li:2014hd}. Although the incident fluence of 1.5 mJ cm$^{-2}$ is relatively strong, the absorption of optical energy in monolayer WSe$_2$ only corresponds to an absorbed fluence of 6.2 $\mu$J cm$^{-2}$, i.e., electron density of 1.3 $\times 10^{13} cm ^{-1}$, calculated based on the monolayer optical reflectivity and complex dielectric constant \cite{Li:2014hd}. The crossing angle between the laser and x-ray pulses is 5$^{\circ}$ in the horizontal plane while the x-ray scattering is in the vertical plane. The temporal resolution is limited by the pump-probe timing jitter on the order of 100 fs. The x-ray diffraction pattern are collected as a function of pump-probe delay in a mode with laser-on and laser-off alternating at 60 Hz repetition rate. The time-dependent reciprocal space maps are analyzed to measure structural dynamics of the monolayer crystals. The sensitivity of the measurement depends on sample quality, experimental geometry and signal-to-noise ratio. Compared to the static measurements, the use of lock-in detection in pump-probe measurements offers enhanced the sensitivity for measuring relative change of the CTRs. In this study, the change of RMSD upon excitation can be measured with an uncertainty of 0.002 $\AA$.

\subsection*{Nonadiabatic quantum molecular dynamics simulations}
Nonadiabatic quantum molecular dynamics (NAQMD) simulations were performed on a simulation cell containing $3\times3$ unit cells of WSe$_2$ monolayer in the 2H crystal structure (9 formula units = 27 atoms) supported on a sapphire (0001) surface. Simulations of free WSe$_2$ monolayers were performed simulation cells containing $3\times3$ unit cells of WSe$_2$ suspended in vacuum. All simulations were performed using the highly-parallelized plane-wave DFT program developed in-house, which can efficiently calculate long-range exact exchange corrections and excited-state forces \cite{Shimojo:2014dt}. 

Electronic structures were described in the framework of density functional theory (DFT) \cite{Hohenberg:1964zz,Kohn:1965ui}, and the generalized gradient approximation was used for the exchange-correlation term \cite{Perdew:1996ug}. Valence wavefunctions are calculated by projector augmented-wave method \cite{Blochl:1994uk}, and projector functions were generated for the 4s, 4p and 4d states of Selenium, 5d, 6s and 6p states of Tungsten, 3s, 3p and 3d states of Aluminium, and 2s and 2p states of Oxygen respectively. The DFT-D method was used to approximate van der Waals interactions in the system \cite{Grimme:2006fc}. The electronic pseudo-wave functions and pseudo-charge density were expanded by plane waves with cut-off energies of 30 Ry and 250 Ry, respectively. All NAQMD simulations were performed in the NVT ensemble with an external Nose-Hoover thermostat at 300K and equations of motion are integrated with a time-step of 0.97 fs.

\begin{acknowledgements}  
 
I.C.T., Q.Z., K.S., G.C., X. X., H.W. acknowledge the support from the Department of Energy, Office of Science, Office of Basic Energy Sciences, under Contract No. DE-SC0012509. AK, HK, AN, FS, RKK and PV acknowledge support of the Computational Materials Sciences Program funded by the U.S. Department of Energy, Office of Science, Basic Energy Sciences, under Award Number DE-SC0014607. NAQMD simulations were performed at the Argonne Leadership Computing Facility under the DOE INCITE program and at the Center for High Performance Computing of the University of Southern California.Use of the Linac Coherent Light Source (LCLS), SLAC National Accelerator Laboratory, is supported by the U.S. Department of Energy, Office of Science, Office of Basic Energy Sciences under Contract No. DE-AC02-76SF00515. Work at the Advanced Photon Source, Argonne was supported by the U.S. Department of Energy, Office of Science under Grant No. DEAC02-06CH11357.
\end{acknowledgements}

\subsection*{Author contributions} 
I.C.T., H.Z., Q.Z., K.S., E.M., C.N., F.E., D.Z., J.G., M.K., S.S., X.X., A.M. performed the measurements. A.K. H.K., F.S., R.K., R.V.,A.N., S.S., P.D. did first-principle calculations. K.S., G.C., X.X. made samples. I.C.T. and H.W. wrote the manuscript with contributions from all authors. H.W. and X.X. conceived this study.

\subsection*{Competing interests} 
The authors declare no competing interests.

\bibliography{wse2}
\bibliographystyle{naturemag}

\end{document}